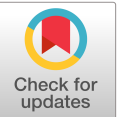

## CLIMATOLOGY

# Uncertainties too large to predict tipping times of major Earth system components from historical data

Maya Ben-Yami[1,2]*, Andreas Morr[1,2], Sebastian Bathiany[1,2], Niklas Boers[1,2,3]*

One way to warn of forthcoming critical transitions in Earth system components is using observations to detect declining system stability. It has also been suggested to extrapolate such stability changes into the future and predict tipping times. Here, we argue that the involved uncertainties are too high to robustly predict tipping times. We raise concerns regarding (i) the modeling assumptions underlying any extrapolation of historical results into the future, (ii) the representativeness of individual Earth system component time series, and (iii) the impact of uncertainties and preprocessing of used observational datasets, with focus on nonstationary observational coverage and gap filling. We explore these uncertainties in general and specifically for the example of the Atlantic Meridional Overturning Circulation. We argue that even under the assumption that a given Earth system component has an approaching tipping point, the uncertainties are too large to reliably estimate tipping times by extrapolating historical information.

## INTRODUCTION

In response to future anthropogenic forcing, some Earth system components might undergo abrupt transitions. These components have come under focus as so-called tipping elements, which are systems that can abruptly change their state under small changes in forcing. This can happen, for example, for systems that exhibit multistability, implying that they could abruptly transition between alternative stable equilibrium states when a critical forcing threshold is passed (1, 2). Such systems include the Amazon rainforest, the Antarctic ice sheets, the Greenland ice sheet (GIS), and the Atlantic Meridional Overturning Circulation (AMOC). Evidence that these Earth system components can indeed abruptly change states comes both from Paleoclimate evidence and from theoretical arguments that transitions can occur under future anthropogenic forcing (2). Transitions of these tipping elements would have severe impacts on climate, ecosystems, and societies from local to regional scales, and their research is thus of high priority. However, both the probability of future tipping and the degree of warming or other forcing factors under which this might happen remain highly uncertain (3, 4). This is in part due to the lack of such abrupt transitions in the recent observational records and in part due to the difficulty of modeling such nonlinear systems using comprehensive coupled climate models. Beyond persisting concerns that these models are biased toward excessive stability (5), they are designed for climate projections with given forcing scenarios, not for predicting individual events in time.

Despite the lack of critical transitions in the observational record, historical observations can still be used to inform us on the changes in stability of Earth system components. When changes in forcing cause multistable systems to approach a transition to a different state, they typically exhibit so-called critical slowing down (CSD), in which their response to perturbations changes in a characteristic manner (6). The most commonly used CSD indicators are the variance and autocorrelation of a time series (7), which increase as the system's stability decreases. In addition, Boers (8) introduced the autoregressive restoring rate λ as a CSD parameter, since λ can be estimated in a way that accounts for nonstationary driving noise. These indicators have been used to identify CSD changes in many systems, including the GIS (9), the AMOC (8, 10), and parts of global vegetation cover (11, 12), in particular the Amazon rainforest (13). As CSD occurs when a system's stability declines on the approach of a critical transition, the identification of these changes can be seen as a warning of such an approaching transitions, and so they are often called early warning signals (EWS) (7).

It may seem natural to take an extra step and use the statistical changes in historical data not only to show a historical and potentially ongoing destabilization but also to extrapolate into the future and predict a tipping time. Although the utility of such predictions, if robust, would be undeniable, the problem lies in the multiple levels of uncertainty inherent to such extrapolations from historical data. In this work, we focus on three types of uncertainties: (i) the modeling assumptions underlying methods for tipping time prediction; (ii) the representativeness of the typically low-dimensional observations, e.g., in terms of fingerprints, of suggested multistable Earth system components that are complex, large, spatially extended, and thus high-dimensional systems; and (iii) the impact of uncertainties and preprocessing on observational datasets, with focus on nonstationary observational coverage and the way gaps are filled.

Below, we will first introduce the three sources of uncertainty in detail and show how they, in general, pose substantial problems for predicting tipping times of any Earth system component. Thereafter, we will go into further detail and exemplify the various difficulties in predicting tipping times from historical data by showing how the different factors influence the predicted tipping time for the AMOC. For the latter, we will focus on a maximum likelihood estimation (MLE) method that was recently introduced by Ditlevsen and Ditlevsen (2023) (14) (hereafter DD23), who applied it to a sea surface temperature (SST)–based fingerprint of the AMOC and predicted that the AMOC would tip around the middle of the 21st century. We show that the described uncertainties are too large to predict a tipping time for the AMOC. Although some of the quantitative results of this work are specific to the AMOC, we show that these types of uncertainties will be present in any attempt to extrapolate a future

[1]Earth System Modelling, School of Engineering and Design, Technical University of Munich, Munich, Germany. [2]Potsdam Institute for Climate Impact Research, Potsdam, Germany. [3]Department of Mathematics and Global Systems Institute, University of Exeter, Exeter, UK.
*Corresponding author. Email: maya.ben-yami@tum.de (M.B.-Y.); boers@pik-potsdam.de (N.B.)

tipping time of proposed Earth system tipping elements from historical data.

## RESULTS
### Sources of uncertainty in tipping time prediction
*Modeling assumptions*

Any method used to predict a tipping time will use past information to extrapolate into the future. To do this, one must make assumptions about the system in question and how it will evolve. Different methods for predicting the tipping time make different assumptions, and in this section, we will show how the methods fail when these assumptions are broken. We will investigate two methods building on conventional CSD-based EWS and one recently introduced maximum likelihood approach (*14*). All three methods assume that the system in question is well-described by a one-dimensional (1D) fold-type normal form bifurcation and that the evolution of the control parameter is linear in time (see Materials and Methods for details):

1) AC(1) extrapolation method: The lag-1 autocorrelation [AC(1)] is a commonly used CSD indicator, as it increases when a system approaches a bifurcation point. For a normal form bifurcation, we can use the known relationship between the AC(1), the local restoring rate γ, and the control parameter α to extrapolate when the system will reach the critical value of α.

2) λ extrapolation method: The AC(1) extrapolation method assumes the noisy disturbances to be stationary in time. However, realistically, these disturbances exhibit nonstationary time correlations. This can be accounted for by basing the extrapolation on the autoregression parameter λ obtained by regressing the increments of the system state $X_{t+\Delta t} - X_t$ onto the states $X_t$ themselves, using a method that accounts for driving red noise of varying correlation strength (*8*). Similarly to the above, in the case of a normal form bifurcation, we can then use the known relationships between λ, γ, and α to extrapolate to the time when α crosses its critical value.

3) MLE method: Third, we investigate the maximum likelihood estimator for fold-bifurcation normal form systems originally proposed in DD23 (*14*). In this approach, the probability density of discrete time increments in the nonlinear fold-bifurcation model is approximated by a discretization scheme. The model parameter choice of maximum probability, based on the observed discrete time increments, is obtained through an optimization routine.

The most fundamental assumption made in all these methods is that the system in question can undergo tipping for a given forcing. However, not all systems can undergo tipping, and as the methods assume tipping, they are susceptible to false positives. We now apply the three methods to time series generated by a linear model without any bifurcation but with an added mean trend, forced with red noise that increases in correlation strength (see Materials and Methods). The first and third method above predict tipping for this system (Fig. 1). For such a linear system, the ideal method would give the tipping time as infinite (i.e., no tipping time). However, the MLE method always predicts a finite tipping time, and the AC(1) extrapolation only gives an infinite tipping time for about a quarter of the cases. The generalized least squares (GLS)–based regression method is designed to account for nonstationary correlated noise, and its results do not indicate a notable decrease in system stability. Despite that, it still gives a finite tipping time for about half of the cases. This is due to the estimation of the slope—since the extrapolation method relies on extending an estimated slope regardless of its significance, the estimated slope will center around 0 but, because of the noise, λ will in practice slightly increase or decrease around 50% of the time (Fig. 1). On the basis of any reasonable statistical significance test, one would rightly conclude that the system in question does not approach a critical transition. Although the risk of false positives is reduced when using the GLS-based regression method to infer λ, one can still not rule out that λ would increase (note that we define λ to be negative so an increase toward zero is interpreted as loss of local stability) for other reasons than the system approaching a bifurcation point; e.g., the local restoring rate of a linear model with only one equilibrium can increase as the result of stretching out the basin of attraction in the vicinity of the equilibrium, resulting in increasing λ although, by construction, the system would not be able to tip.

Even if the system can undergo tipping, many internal and external complexities can cause the above methods to be substantially biased. We examine the methods' performance in the most commonly used model for a system with a bifurcation: the fold normal form bifurcation. In reality, we cannot assume that the fold normal form is applicable to climate tipping elements, as we will discuss specifically for the AMOC in the next section. Nevertheless, here, we show that tipping point prediction can fail even when the assumption of this simple model holds.

The normal form of a fold-type bifurcation is given by

$$\frac{dx}{dt} = -b(x(t)^2 - \alpha) \quad (1)$$

where $x$ is the system state, $b$ is a timescale parameter, and $\alpha$ is the bifurcation parameter of the system. The stable equilibrium $x^*(\alpha) = \sqrt{\alpha}$ is characterized by a linear restoring rate of $\gamma = 2b\sqrt{\alpha}$. Under the assumption that the system is forced by white noise, one can devise the model-specific extrapolation of the bifurcation parameter by examining $\log[AC(1)]^2 \sim \alpha$ and $\log(\lambda + 1)^2 \sim \alpha$. If this bifurcation parameter evolves linearly, then this gives a more adequate estimation of the tipping time than linearly extrapolating the indicators themselves (see Materials and Methods). The DD23 MLE method was also developed specifically for this model structure. In Fig. 2A, we give the distributions obtained by applying the different methods to $10^4$ sample time series of length 2000 time units each, all from a fold normal form with white noise and linear forcing. The method proposed by DD23 performs best, which is to be expected, as that method makes use of the known model structure in a statistically optimal way. The two methods relying on extrapolating trends in CSD indicators are biased toward too early tipping times and exhibit large spread. This is mostly due to the propagation of estimation errors of AC(1) and $\lambda$ through the inversion $\log(\cdot)^2$ with respect to their relation to $\alpha$ and through the subsequent linear regression. While this estimation error could be avoided by fitting the exponential directly, such a fit comes with its own substantial errors.

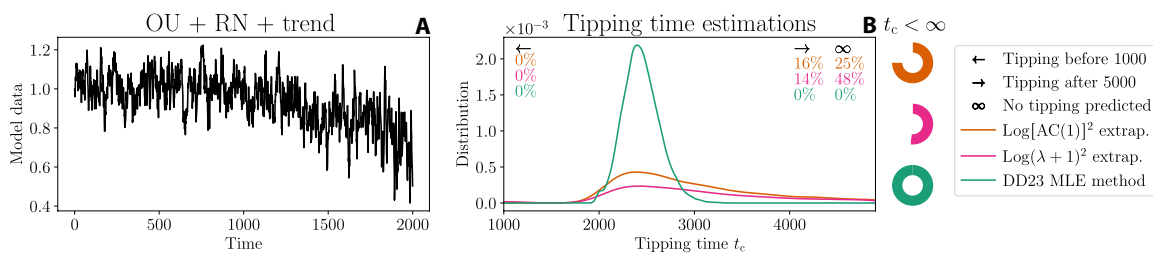

**Fig. 1. Tipping time estimation for data stemming from a linear model with no possibility of tipping.** (**A**) A time series of an Ornstein-Uhlenbeck (OU) process driven by nonstationary red noise (RN) is shown. A nonlinear decreasing mean trend is added to the resulting time series realization. Such low-frequency variability is commonly observed in natural systems and need not indicate an approach toward a tipping point. It could likewise be due to an only partly observed oscillation. (**B**) The three methods for estimating the tipping time are used on $10^4$ trajectories of the linear system: AC(1) extrapolation (orange), λ extrapolation (pink), and the MLE method (turquoise). The distributions are estimated using standard Gaussian kernel density estimation and then scaled to the fraction of times estimated between 1000 and 5000. Numbers in the figure corners show the percentages of estimates outside the figure range and at infinity. Although no tipping is possible, the method relying on the AC(1) and the MLE approach purport such an existence in a large fraction of cases, as shown in the pie charts on the right. The method relying on a GLS model comes close to the ideally expected performance of indicating tipping in 50% of the cases, which is due to an underlying constant stability being symmetrically estimated (see the text).

As already seen above (Fig. 1), assuming that uncorrelated noise is driving the system (which in realistic situations is indeed rarely justified) can lead to complications when attempting to estimate the tipping time. Accordingly, the first and third methods are biased toward too early tipping times when applied to data from a fold normal form system driven by red noise with increasing correlation strength (Fig. 2B). For the AC(1) extrapolation, this bias adds on top of the estimation error described above. In contrast, for the $\lambda$ extrapolation method using the GLS model, the nonstationary red noise does not induce an additional bias on top of the estimation bias. Last, for the MLE method, there is a clear shift toward too early tipping times compared to the white noise case.

### Stationarity of past trends

In addition to making assumptions about the underlying dynamical model, to predict a future tipping time, we also have to assume how the forcing of the system will change in the future. In particular, all of the methods above rely on the fact that the forcing will continue to evolve in the same way it has in the past.

It is typically not easy to identify the exact climate variable that acts as the primary forcing parameter for a given climate tipping element. For example, while one might assume that global mean temperature (GMT) might be appropriate for the Greenland and Antarctic ice sheets, all the involved processes and the different latitudinal temperature distributions indicate that this would likely be an oversimplification (15). Similarly, for the Amazon rainforest, one might think that mean annual precipitation is the likely forcing parameter, but the rainfall seasonality and dry season length play important roles as well (16, 17). For the AMOC, the effective freshwater flux in the north Atlantic could be considered the key control parameter, but again, given that this is a spatially extended system with shifting regions of deep water formation, it is unclear how to define an accurate control parameter in detail (*18*). For any of these systems, there is no evidence that their control parameters, even if known, would be linearly dependent on GMT. Furthermore, although the logarithm of the atmospheric $CO_2$ concentration (and hence its radiative forcing) has increased linearly since 1850, it is unlikely that this linear increase will continue. The only scenario under which log($CO_2$) would continue to increase linearly is the extreme of Shared Socioeconomic Pathway 5.85 (SSP5.85) (fig. S1), a scenario in which there is not only no climate mitigation but also a rapidly growing fossil fuel–based economy (*19*).

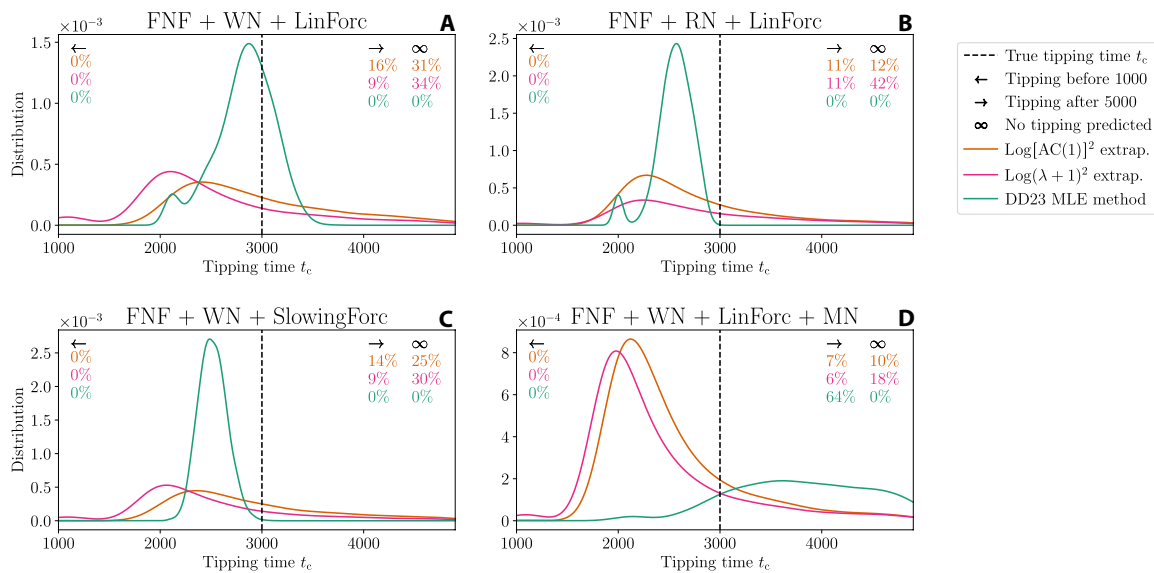

**Fig. 2. Tipping time estimation for data stemming from different variations of the fold-bifurcation normal form (FNF) model.** (**A**) The distributions are derived from $10^4$ model trajectories (see Materials and Methods for all model equations). The three methods used are AC(1) extrapolation (orange), $\lambda$ extrapolation (pink), and the MLE method (turquoise). In (A), the model is driven by white noise (WN) and a linearly evolving forcing $\alpha$. This constitutes the case for which all considered estimation methods were designed. For the MLE method, the accumulation of estimates at time 2000 is rooted in the method's inherent optimization constraints. A tipping time earlier than the time series end at time 2000 is not possible in this framework, and the optimization terminates at this value. (**B**) Results for a model driven by nonstationary red noise (RN) with increasing correlation strength and a linearly increasing forcing. The model investigated in (**C**) exhibits a nonlinear trend of the bifurcation parameter, which decelerates as time progresses. The time series underlying the results of (**D**) are identical to those of (A), barring an added white measurement noise (MN), whose strength decreases as time progresses, mimicking reduced observational uncertainties over time. Note that the two CSD-based methods remain biased toward too early tipping time estimates, while the MLE methods is, in this case, biased toward too late estimates. All distributions are estimated using Gaussian kernel density estimation and scaled to the fraction of times estimated between 1000 and 5000. Numbers in the figure corners show the percentage of estimates outside the figure range and at infinity.

In addition, while anthropogenic global warming is usually assumed to be driving climate systems toward a tipping point, it is not the only substantial factor altering their conditions. For example, deforestation can be a large factor in destabilizing tropical rainforests (*20*–*22*). Such time-varying influences cannot be represented by a linearly evolving control parameter. Thus, while the conceptual view of only one external control parameter evolving linearly toward a critical value is sometimes sufficient to describe the historical time evolution of a given climate system, we argue that it is too simplistic to allow for an estimate of the tipping time.

The three extrapolation methods discussed above assume a linear trend in the bifurcation parameter α. They therefore produce biased results when applied to time series of a model whose bifurcation parameter changes nonlinearly, e.g., in a decelerating manner. Figure 2C shows the corresponding estimation results. In principle, any extent of misinterpretation of observations is possible if the underlying evolution of a system toward a tipping point cannot be assumed to be stationary.

### Representativeness of measurements for underlying dynamics

The above modeling uncertainties arise when one assumes that the time series used are a direct representation of the climate system in question. In practice, that is rarely the case. Tipping elements in the climate are complex, spatially extended systems with many degrees of freedom, and it is always a crude simplification to describe their dynamics by a 1D observational time series. Moreover, it is often unclear whether the relevant dynamical properties of the system in question are captured well by the available measurements. For example, the Amazon rainforest is observed using different remotely sensed vegetation indices, but it remains hard to tell which one is appropriate for understanding its stability (*11*, *23*). The problem is further complicated by the relatively short time span of many climate observations. Tipping elements such as the polar ice sheets or the AMOC evolve over long timescales, and so, time series of a hundred years or more would be necessary to understand and predict their dynamics. When there are no direct measurements on such long timescales, studies use different proxies ("fingerprints") for the systems. These fingerprints are climate variables that have some physical connection to the system of interest and are thus, to some extent, correlated with the changes of the system. For example, ice core–derived Greenland melt rates have been used as a fingerprint for ice sheet height and SST patterns as a fingerprint for the AMOC streamfunction strength (*9*, *24*). Similarly, satellite-derived vegetation indices should be interpreted as—uncertain and potentially biased—fingerprints of the actual vegetation dynamics (*11*, *23*). Fingerprints that are useful for understanding the evolution of the mean trend are not necessarily useful for predicting tipping times. In conclusion, there are often no fingerprints available with the precision required for predicting tipping times, as we will discuss for the AMOC in the next section.

### Effect of dataset preprocessing and underlying uncertainties / nonstationary coverage

In addition to the uncertainties arising from the modeling approach and the choice of fingerprint, there are also substantial uncertainties in CSD indicators that originate from the dataset preprocessing steps and the nonstationarity of observational data coverage (*23*, *25*, *26*). The number of available climate observations has grown exponentially since 1850 (*27*, *28*), especially with the increase of remote sensing measurements since the 1970s. In earlier years, these measurements are often concentrated in a small number of areas on the globe, resulting in uneven and sparse global coverage until at least the mid-20th century (*28*). Therefore, to produce globally complete datasets, researchers often merge a number of different instrumental records and fill in the gaps in data using a processing procedure. The bias correction for the different measurements and the infilling methods often prioritize the accuracy of the mean trend over the accuracy of the higher-order statistics [see, for example, (*23*, *29*–*32*)].

This can cause problems for the detection of CSD (*23*, *25*, *26*, *33*) and especially for the calculation of the future tipping time. Any calculation of the future tipping time is strongly reliant on the time evolution of the data's higher-order statistics, and the dataset preprocessing can induce artificial trends in these statistics. To this point, we show the effect of adding white measurement noise of a decreasing amplitude to the synthetic model data of the fold normal form and estimate tipping times using the introduced methods. Figure 2D shows how a small amount of observational uncertainty can incur substantial changes in the estimations. One real-world example of such an effect is the merging of multiple satellite signals. The change in signal-to-noise ratio from one satellite to the next can cause an artificial increase in autocorrelation (*23*). Another issue emerges when missing data are infilled with some sort of principal components analysis, in which case there will be more artificial smoothing in earlier times due to the lack of data, and so quantities like the variance could increase artificially (*26*, *29*). In the next section, we will discuss such dataset uncertainties in detail for the AMOC SST fingerprints.

## Uncertainties in predicting the AMOC tipping time

We now address these uncertainties for the specific case of predicting the tipping time of a system by applying DD23's MLE-based method to SST-based fingerprints of the AMOC (*14*). We choose to use DD23's methodology because among the three discussed methods, it performs best for a fold normal form with white noise (Fig. 2A) and was designed for AMOC tipping time prediction.

### Modeling assumptions

There has long been a discussion about whether the AMOC, when investigated as a complex system under external forcing of, e.g., GMT (or better, regional freshwater forcing), exhibits multiple stable states (*34*–*36*). Transitions between such stable states could be bifurcation induced and thus abrupt and irreversible. The so-called fold bifurcation constitutes a minimal example of such behavior. For instance, the conceptual Stommel model of the AMOC features a fold bifurcation (*34*). Taking this reasoning another step further, the application of the MLE method assumes that the 1D observable of AMOC strength is well represented by the following normal form model

$$dX_t = -b((X_t - m)^2 - \alpha)dt + \sqrt{b}\sigma dW_t \quad (2)$$

where $X_t$ is the system state at time $t$, α is the external control parameter, $b$ is a timescale parameter, $m$ is a translation parameter, and the white noise term $\sigma dW_t$ represents noisy perturbations within the system. A similar simplification has been applied to more complex AMOC models in multiple studies (*37*, *38*) and is justified by arguing that all fold bifurcations are topologically equivalent to this model. This, however, is only true locally, in potentially very close proximity to the bifurcation point, while the proposed estimation method banks on the assumption that it would hold in arbitrary distance to the bifurcation point. On the basis of the arguments brought forth, we do not see a direct constraint on the dynamics away from the tipping point. When applying the MLE method to data of the closely related 2D Stommel-Cessi AMOC model (*39*), one obtains a considerable bias in tipping time estimates toward earlier times (Fig. 3A). This should be seen as an indication that even models with an underlying fold bifurcation structure, yet not following the very specific normal form model equation above, produce time series which, when applying the MLE method, yield biased tipping time estimates.

The AMOC exhibits pronounced decadal variability (*40*). Before the commencing of the destabilisation at time $t_0$, the AMOC is assumed to resemble paths of a stationary stochastic process $X$ defined by

$$dX_t = -2b\sqrt{\alpha}(X_t - m)dt + \sqrt{b}\sigma dW_t \qquad (3)$$

When applying their MLE method to the AMOC, DD23 estimates a value of $2b\sqrt{\alpha} \approx 3.1$ [year$^{-1}$], corresponding to a characteristic correlation time of 0.32 [year]. In contrast, frequency spectra of AMOC evolutions in general circulation models show strongest variability between 5 and 100 years [e.g., figure 6 in (*41*)]. Such pronounced additional variability on long timescales is not captured by the above Ornstein-Uhlenbeck model. Internal variability independent of the model noise will thus cause large excursions from the transient mean. The proposed method is not equipped to incorporate the impact of these excursions on the estimated tipping time, since they may be misinterpreted as trends toward a tipping point. This exposes the estimation method to risks of false alarms of a similar nature as in Fig. 1. In addition, recent application of DD23's MLE method to AMOC tipping in a complex climate model (*42*) has shown that the tipping time prediction is very sensitive to the time interval analyzed due to the decadal variability of the AMOC, and most 150-year windows cannot accurately estimate the tipping time.

Moreover, for quantitative extrapolations of tipping time, any simplifying assumptions on the driving noise would need to be carefully checked. Since disturbances to the equilibrium state are themselves of atmospheric and oceanic origin, time correlation of the noise should be taken into consideration, e.g., via a red noise model (*8*). Nonstationary red noise present in the system can incur substantial biases in the estimation of the tipping time and even result in false alarms of an approaching bifurcation (as seen in Figs. 1 and 2B).

### Assumptions on future AMOC forcing

Previously, we discussed the fact that not only can we not assume the future evolution of the forcing of climate tipping elements to be known, we also cannot assume that the forcing evolved linearly in the past. This is also true for the AMOC: Several studies show that radiative anomalies due to aerosol pollution likely attenuated the AMOC weakening of the past decades (*43*, *44*), and such changes cannot be modeled with a linearly changing control parameter. Moreover, the GMT forcing itself influences the AMOC due to many different nonlinear mechanisms, e.g., via thermal expansion, a strengthening hydrological cycle, as well as sea ice and ice sheet melt (with the influence of the latter in the historical period still under debate) (*45*, *46*). The effective freshwater flux might serve as a better forcing parameter (*35*, *47*), but we do not have long-term measurements of this parameter, and there is evidence that it does not linearly depend on GMT; e.g., Greenland runoff increases nonlinearly over time (*48*, *49*).

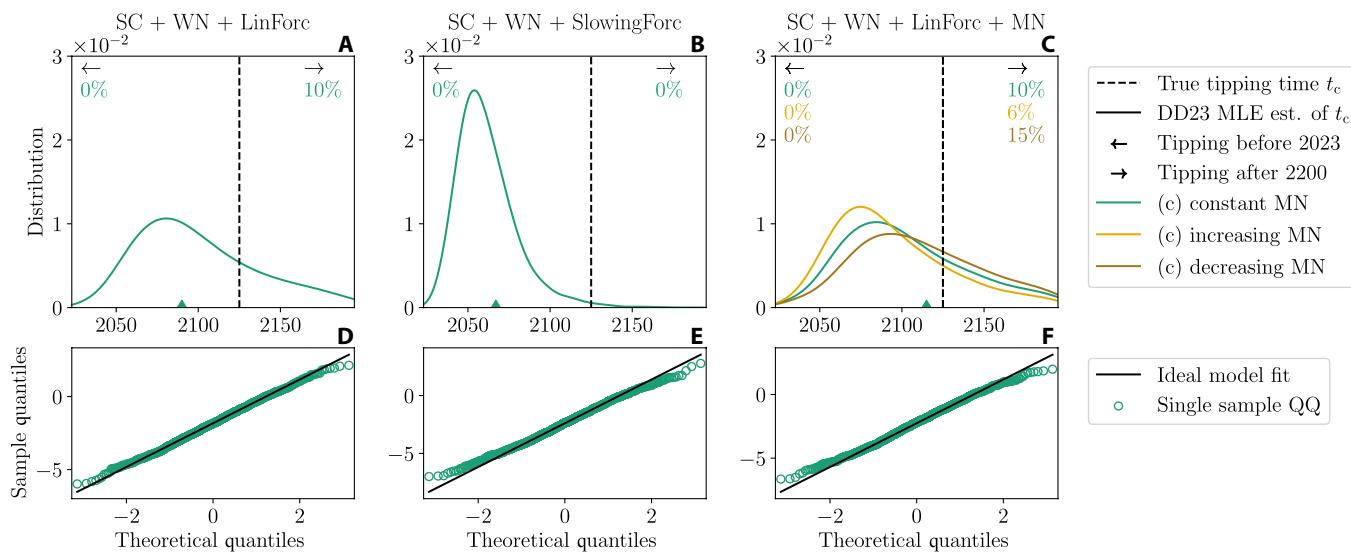

**Fig. 3. Tipping time estimation for data stemming from variations of the conceptual Stommel-Cessi (SC) model.** The distributions depict the estimations of the tipping time for $10^4$ model runs (see Materials and Methods for model equations). (**A**) Tipping times estimated from data obtained from the 2D Stommel-Cessi model with a linear forcing parameter. (**B**) Same as (A) but using a nonlinear forcing which decelerates over the integration time span. This is a model setting with practical relevance, as the anthropogenic changes forcing the AMOC cannot be assumed to be constant. (**C**) Same setting as (A) but with white measurement noise added, with constant (green) increasing (light brown) and decreasing (dark brown) amplitude. As above, distributions are estimated with Gaussian kernel density estimation and then scaled to the fraction of times estimated between 1000 and 5000. Numbers in the figure corners show the percentages of estimates outside the figure range and at infinity. The tipping time estimates in each panel are biased because the data does not stem from the exact intended model expected by the MLE method of DD23. (**D** to **F**) The Quantile-Quantile (QQ) plots beneath each panel give the model fit of the derived maximum likelihood model to the data of one sample. The tipping time estimate of the respective sample is indicated by the triangle in the top panel. A comparison of the QQ plots suggests that time series stemming from the 2D Stommel-Cessi model are similarly well-modeled by the proposed fold normal form model with white noise forcing as the AMOC time series of DD23 [figure 6F in (*14*)].

### Representativeness of the subpolar gyre SSTs for AMOC

Because of the lack of long-term observations, various fingerprints have been proposed (*50*). The most commonly used fingerprint for the AMOC is based on SSTs in the subpolar gyre (SPG), and a modified version of this fingerprint is used by DD23 to predict AMOC tipping. This fingerprint is based on the assumption that the so-called warming hole in the North Atlantic, an area which is cooling as opposed to the global warming trend detected essentially everywhere else, is caused by a weakening of the AMOC (*51*–*54*). The classical fingerprint is defined as the SSTs averaged over the SPG area minus the global SST mean (*24*, *51*) and has often been referred to as the "SPG Index." However, that term is also used in the literature to describe indices related to the characteristics of the SPG circulation (*55*). In this work, we therefore call this AMOC fingerprint the SPG-based AMOC index.

This index has been supported by two lines of evidence. First, across models, the historical trends in the SPG-based AMOC index in CMIP6 models correlate with the trends in the AMOC streamfunction (at various latitudes), in the sense that models with higher AMOC strength trend also have higher SPG-based AMOC index trends (*24*, *44*). Second, the SPG-based AMOC index time series itself is correlated with the AMOC streamfunction time series at various lag times (depending on the study either the maximum of the streamfunction is taken or its value at different latitudes) (*51*, *56*–*58*). However, both these correlations have been shown to be highly nonstationary and are sensitive to the time period, to the forcing scenario and to the underlying processes (*56*, *58*). This is likely due to the fact that the warming hole is not driven solely by the AMOC but is a result of both changes in ocean heat transport and changes in atmospheric forcing (*59*–*63*). This partial connection of the SPG to the AMOC is supported by recent studies using the Overturning in the Subpolar North Atlantic Program, which have shown that the Labrador Sea and the SPG play a smaller role in North-Atlantic deep water formation than previously thought (*64*, *65*).

The nonstationarity of the correlation between AMOC streamfunction and the SPG-based AMOC index does not imply that this index is not useful for studying the stability of the AMOC, as the SPG still plays a crucial role in the AMOC and would thus be sensitive to its stability changes (*26*, *45*, *66*, *67*). Signs of CSD in the SPG region thus still likely indicate a destabilization of the AMOC. However, the nonstationarity does reduce the fingerprint's usefulness for exact predictions of tipping times. This is particularly true due to the lack of agreement over the precise nature of this nonstationarity, which means that we have no way of accounting for it when using the fingerprint. We believe that for predictive purposes, including those based on extrapolation, it is problematic to fit a simple bifurcation model representing the AMOC to a fingerprint whose correlation with the AMOC changes over the time period under consideration.

To obtain a better representation of the AMOC, different proposed fingerprints should be compared. The uncertainty in fitting a model to the SPG-based AMOC index alone can then be inferred by comparing the results of the CSD analysis and extrapolation for the different fingerprints. There is a long list of identified AMOC fingerprints in the literature, and many of them as robust and commonly used as the SPG-based AMOC index (*50*, *57*). When one applies DD23's MLE method to one of these other fingerprints, the so-called dipole fingerprint (*68*), the estimated tipping time varies considerably and sometimes even goes to infinity (Fig. 4 and tables S1 and S2). Since there is now no consensus on which of these fingerprints better represents the AMOC, the range of estimated tipping times highlights substantial uncertainty in such estimations.

It should also be noted that there is growing evidence supporting the SPG as a potential tipping element separate from the AMOC (*1*, *69*, *70*). Although an SPG collapse occurs only in some coupled climate models under future warming scenarios, these models are among the best in representing the stratification in the SPG (*69*, *70*). We cannot, therefore, disregard the possibility that CSD in the SPG-based AMOC index is in reality an indication of an approaching SPG tipping point and not an AMOC tipping point. The only way to avoid this uncertainty is to include additional AMOC fingerprints which do not rely on SPG SSTs (*8*).

### Uncertainties from the SST datasets

We have shown in a conceptual example how simple forms of measurement noise can cause complications for the estimation of the tipping time (see Figs. 2D and 3C). For the prediction of an AMOC tipping time, the main source of uncertainty arising from the SST datasets is caused by the infilling methods. For example, in their study, DD23 uses the HadISST1 SST dataset, which has been infilled using reduced space optimal interpolation (RSOI) (*29*). RSOI uses a set of global empirical orthogonal functions (EOFs) and includes regularizing terms when fitting the EOFs to the data. This is done to avoid spurious large amplitudes in data-scarce regions and times but means that the fit tends to the zero anomaly where there is no information. Although noninterpolated in situ data are subsequently added to the RSOI reconstruction, this only improves the variance where there is enough data—in data-scarce times and regions, the variability is damped by RSOI. Together with other steps of the preprocessing, this causes the variance in HadISST1 to artificially increase [see (*26*, *29*)].

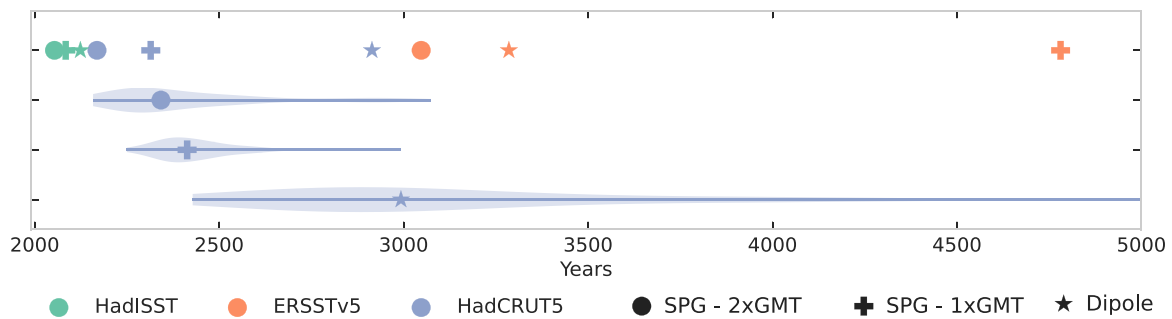

**Fig. 4. Range of tipping times.** Tipping times estimated using DD23's MLE method. The best estimate of the tipping time is calculated for the classical SPG-based AMOC index (plus), the fingerprint used by DD23 (circle), and the Dipole index (star). We use three different observational SST datasets for this analysis: HadISST1 (turqoise), ERSSTv5 (orange), and HadCRUT5 (blue). In addition, the blue violins show the tipping times for each of the 200-member uncertainty ensemble of HadCRUT5. (See fig. S2 for a version of the tipping time calculation applying an additional penalization; cf. section S2 of DD23 for details). The plotted values can be found in tables S1 and S2.

To highlight the effect of dataset processing methods on the tipping time calculation, we use DD23's MLE method to calculate tipping times for the AMOC, using three different datasets: the previously mentioned HadISST1 (*29*), HadCRUT5 (*30*), which uses a Gaussian process–based statistical method for infilling, and ERSSTv5 (*32*), which uses empirical orthogonal teleconnections for infilling. All of these dataset methods result in different variance and autocorrelation time series (see Fig. 5), as does the noninfilled HadSST4 (*28*). The variance is especially affected by the various preprocessing methods of the different datasets—only in HadISST1 does the variance increase over the whole time period—and as noted above, this increase is at least partly artificial. It is therefore not possible to determine the actual variance trend of north Atlantic SSTs before the 1970s. While the autocorrelation and the restoring rate are arguably still functional indicators given the dataset properties [see (*26*)], DD23's MLE method relies on the variance and does not take these uncertainties, the nonstationary observational coverage and the different gap filling procedures into account.

When applying DD23's MLE method to their version of the AMOC fingerprint but calculated from alternative SST datasets, we obtain tipping times ranging from the 2000s for HadISST to the 3000s for ERSSTv5. If this analysis is extended to different AMOC fingerprints (see the previous subsection), then the tipping times range from the 2000s to beyond the year 4700 for ERSSTv5 (table S2). Last, if we apply the method to HadCRUT5's 200-member uncertainty ensemble, we get multimillennial uncertainty ranges with, for some cases, almost a quarter of the tipping times going to infinity (table S1). This shows that the fingerprint definition and the dataset choice can cause huge uncertainties.

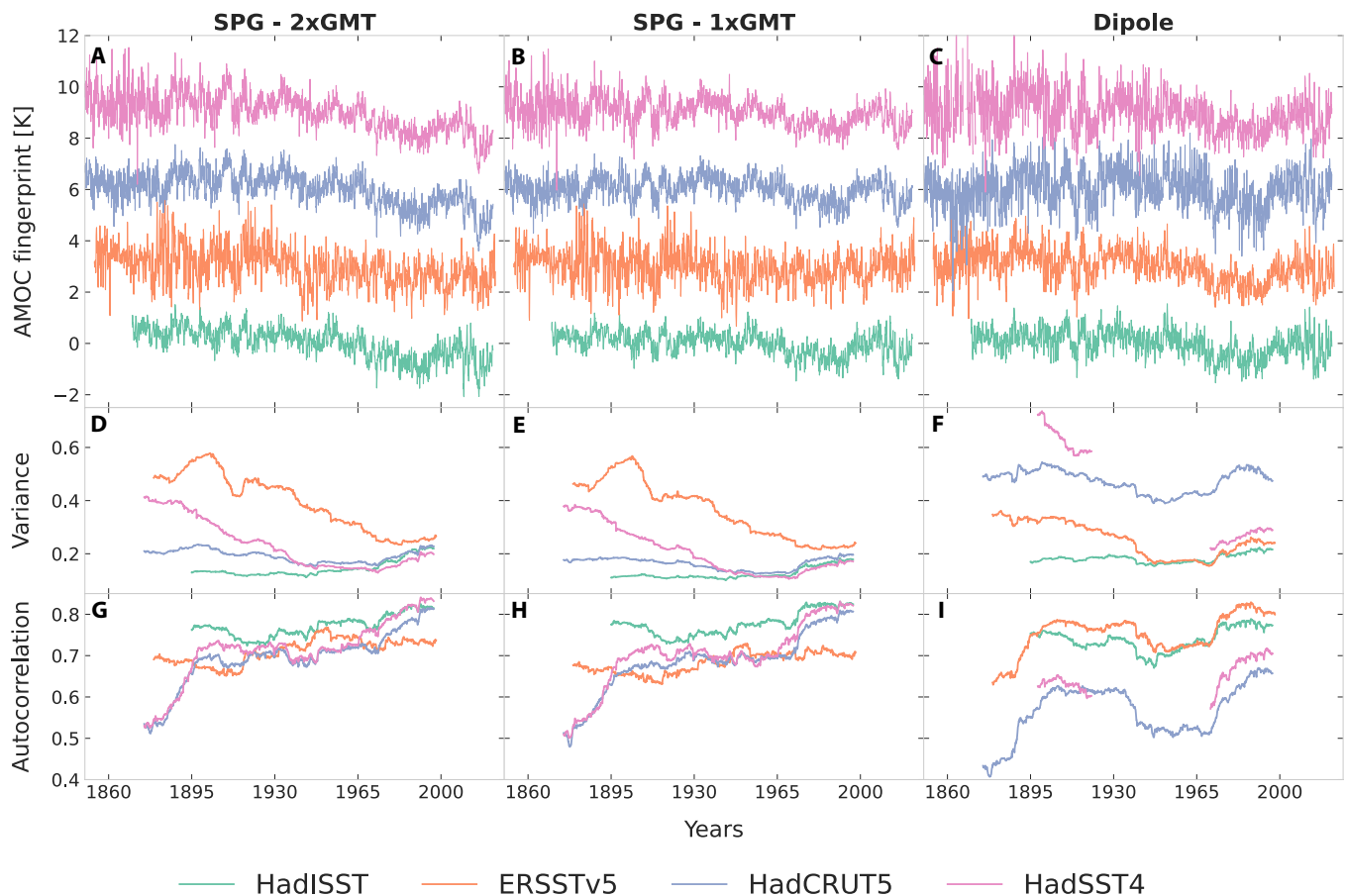

**Fig. 5. Variance and autocorrelation for different SST datasets and AMOC fingerprints.** The rows from top to bottom show the monthly AMOC fingerprints (**A** to **C**), variance (**D** to **F**), and autocorrelation (**G** to **I**). The columns from left to right show the values for the fingerprint from DD23 (left), the classical fingerprint from Caesar *et al*. (*24*) (middle, SPG-based AMOC index), and the AMOC dipole fingerprint (*50*) (right). The dipole is defined as averaged SSTs in 45° to 80°N, 70°W to 30°E minus SSTs in 0° to 45°S, 70°W to 30°E. The time series are shown for four different datasets: HadISST1 (turqoise), ERSSTv5 (orange), HadCRUT5 (blue), and HadSST4 (pink). In (A) to (C), the AMOC fingerprints are offset by 3 K from each other for better visibility. All CSD indicators are computed using a window size of 50 years. Note that the variance shows overall decreases in most cases, partly due to the nonstationary data coverage (*26*). In addition, note that the calculation of the SPG-2xGMT fingerprint in this work is slightly different from in DD23 (see Materials and Methods).

## DISCUSSION

Regardless of the Earth system component under consideration, it is inevitable that at least some of the assumptions discussed above will be broken in reality and that the types of uncertainties addressed in this work will indeed arise when attempting to extrapolate tipping time from past historical data. First, simplified modeling assumptions will almost always be necessary for extrapolation, since the future behavior of the system will be different depending on the governing dynamics. Past data can inform us about the relevant model, but typically, many different models can match the data, as seen above. Second, the problem of finding a time series that accurately represents the dynamics of the system is common to all tipping elements. Last, the problems caused by nonstationary data coverage and data processing methods described above are unavoidable, since data processing is always necessary to assimilate and calibrate observations and proxies, especially for longer records.

However, it is important to emphasize that the criticisms in this work apply to attempts to predict the exact tipping time of tipping elements such as the AMOC, based on extrapolating from uncertain data. CSD detection in terms of trends in robust indicators such as the autocorrelation or the restoring rate is much less sensitive to the discussed uncertainties, whereas the variance should indeed be used with caution (*8*, *25*, *26*). In principle, CSD is applicable to any sort of dynamical system that is approaching a transition induced by a codimension one bifurcation. In addition, fingerprints that are not an exact representation of a dynamical system will still show CSD as long as the stability of the subsystem they represent is connected to the stability of the overall system. Crucially, the uncertainties presented in this work can be taken into account by using multiple different fingerprints and propagating the dataset uncertainties to the CSD analysis. Such an analysis has already been applied to the AMOC by Ben-Yami *et al.* (*26*), who found that CSD in AMOC fingerprints in terms of a restoring rate tending toward zero, is still significant although the trends in the CSD indicators have a large spread. Taking into account the same observational uncertainty spread for the tipping time, however, gives time ranges from 2050 to infinity, practically making this prediction noninformative. This is because tipping time prediction is not only more sensitive to uncertainties but also relies on more assumptions and in particular presumes that there will be a tipping time in the future. Therefore, for the prediction method to be useful, it needs to narrow down the future range of tipping times to an informative range. In contrast, a detection of CSD does not make statements about future tipping, only about the fact that the system is now less stable than it was in the past.

Although the example we put most focus on was the predicted tipping time of the AMOC, the work of Boers and Rypdal (2021) (*9*) (hereafter BR21) could also be interpreted as a prediction of tipping time for the central-western part of the GIS. The same kinds of uncertainties apply to that work. BR21 derive the potential landscape of the ice sheet height by fitting a previously introduced nonlinear model (*71*), which makes many assumptions. The change in ice sheet height is calculated from average annual melt rates obtained from three ice cores in central-western Greenland, and in addition to uncertainties in the underlying data that are difficult to quantify, it is highly uncertain how well this location represents the whole ice sheet (*49*). A potential tipping time could be identified with the bifurcation point of the fitted model (red vertical dashed line in figure 3 of BR21). However, the uncertainties from the assumptions of the simplified model and the reconstruction uncertainties imply that the estimated bifurcation point should not be understood as an estimate of the actual critical threshold and should certainly not be translated into a tipping time (*9*).

We have discussed multiple sources of uncertainty in the prediction of future tipping times of Earth system components. These uncertainties are as follows:

1) The modeling assumptions underlying the methods for tipping time estimation

2) The viability of extrapolating past forcing trends into the future

3) The reliability of using indirect fingerprints to predict tipping times of climate tipping elements

4) The uncertainties that arise from the bias and preprocessing in observational datasets with measurement uncertainties and gaps

The latter two points above may be addressed in time by improved Earth system observations and waiting (possibly for hundreds of years, depending on the characteristic timescale of the system) until sufficiently long records are available. Regarding the first point, it is unclear whether the highly nonlinear and complex dynamics governing the proposed tipping elements will ever be reliably modeled at the accuracy needed for tipping time prediction. Last, regarding the second point, it will never be possible to know the change in future forcing, so any extrapolation will always be uncertain as it would assume a specific future scenario.

In addition, we have described in detail how these uncertainties manifest for the specific example of predicting a future AMOC tipping time, using the MLE method introduced by DD23 (*14*):

1) The modeling assumptions underlying DD23's MLE method for tipping time predictions are too simple and do not necessarily hold for the AMOC. We have shown that breaking these assumptions by, e.g., changing the dynamical model for the AMOC or the model for the forcing introduces large biases in the tipping time estimation (Fig. 4). In particular, their method also predicts a tipping time for a linear model that cannot tip, if forced by red noise with increasing correlation strength.

2) The connection of the SPG-based AMOC fingerprint (computed from SSTs) to the AMOC is uncertain and nonstationary and therefore is problematic for exact predictions of tipping times. Using different SST fingerprints with the HadISST dataset can change the predicted tipping time by 70 years (table S2).

3) The inherent uncertainties of SST datasets and the preprocessing methods used to fill in missing data can be nonstationary and thus affect higher-order statistics such as the variance or autocorrelation. In particular, the HadISST dataset used by DD23 is known to have an artificial variance increase. Using different SST datasets and their uncertainty ensembles, the tipping time varies by thousands of years (tables S1 and S2).

In the foreseeable future, points 2 and 3 will essentially form impassable barriers to predicting the time of a future AMOC collapse from historical data. The available data are simply not accurate or precise enough to make such an extrapolation.

In conclusion, we showed that the uncertainties discussed in this work are too large to allow for reliable estimates of the tipping time of major Earth system tipping elements, including the AMOC, the polar ice sheets, or tropical rainforests, based on extrapolating results from historical data. We emphasize that these uncertainties, originating from underlying modeling or mechanistic assumptions as well as from the used empirical data, need to be taken into account and propagated thoroughly before attempting to estimate a future tipping time of any potential Earth system tipping element.

## MATERIALS AND METHODS
### AMOC fingerprints

For each of the four used SST datasets, we compute three different SST-based fingerprints of the AMOC. First, the index introduced by DD23, which is obtained by averaging SSTs over the SPG region and then subtracting twice the global mean SSTs. Here, the SPG region is defined as in (*24*). Second, the original version of this index, introduced by Caesar *et al.* (*24*), where the global mean SSTs are only subtracted once. Third, we use the so-called dipole fingerprint, which is obtained by subtracting average SSTs of a large region in the southern hemisphere Atlantic ocean (0° to 45°S, 70°W to 30°E) from average SSTs in a large region in the northern hemisphere Atlantic (45° to 80°N, 70°W to 30°E). For computing the spatial averages for the HadISST data, we mask out all values of grid cells covered by sea ice following (*8*, *24*) and also use a weighted mean to account for the dependence of the grid cell size on the latitude.

### Formulae for conceptual models

Here, we give the formulae of all models which have been integrated to obtain time series data for the subsequent analyses in Figs. 1 to 3. The abstract models of the first two figures were integrated for a total simulation time span of 2000 time units and sampled at time step 1. For the first 1000 time units, the dynamics are held constant, i.e., no evolution in the bifurcation parameter or the noise is present. At time $t_0 = 1000$, these nonstationarities commence. They are aimed toward a horizon of $t_c = 3000$. The ramp duration is thus $\tau_r = 2000$. For the application in the AMOC setting, we emulate AMOC time series data from the year 1870 to 2021 and generate these at a monthly time step, i.e., 1/12 years. We emphasize, however, that there is no seasonality in these simulated time series. The ramp parameters are $t_0 = 1924$, $t_c = 2125$, and thus $\tau_r = 201$.

The linear model of Fig. 1 is an Ornstein-Uhlenbeck process driven by another Ornstein-Uhlenbeck process. After the integration, a mean trend is added to the dynamics.

$$dX_t = -\gamma X_t dt + \kappa U_t dt \quad (4)$$

$$dU_t = -\frac{1}{\tau^{\text{noise}}(t)} U_t dt + dW_t \quad (5)$$

$$\tau_{\text{noise}}(t) = \tau_0^{\text{noise}}(1 - \Theta[t-t_0](t-t_0)/\tau_r) + \tau_{t_c}^{\text{noise}} \Theta[t-t_0](t-t_0)/\tau_r \quad (6)$$

$$x^{\text{trend}}(t) = \sqrt{1 - \Theta[t-t_0](t-t_0)/\tau_r} \quad (7)$$

where $dW_t$ is the white noise forcing, making the Ornstein-Uhlenbeck process $U$ a red noise forcing. $\Theta$ is the Heaviside function, $\gamma = 0.2$ is the linear restoring rate, $\kappa = 0.023$ is the noise strength, and $\tau_0^{\text{noise}} = 1/3$ and $\tau_{t_c}^{\text{noise}} = 2$ determine the linear evolution of the noise correlation. Further, $\tau_r$ determines both the ramp of the noise correlation and of the added trend. The original fold bifurcation normal form model with white noise forcing is given by the following equation for $dX_t$. The equations for $\alpha(t)$ define different forcings: linear forcing [$\alpha^{\text{lin}}(t)$] or decelerating forcing [$\alpha^{\text{slow}}(t)$]

$$dX_t = -b(X_t^2 + \alpha(t))dt + \sqrt{b}\sigma dW_t \quad (8)$$

$$\alpha^{\text{lin}}(t) = \alpha_0(1 - \Theta[t-t_0](t-t_0)/\tau_r) \quad (9)$$

$$\alpha^{\text{slow}}(t) = \alpha_0(1 - \Theta[t-t_0](t-t_0)/\tau_r)^{1.5} \quad (10)$$

Here, $b = 0.1$ is a timescale parameter, $\sigma = 0.05$ is the amplitude of the noise, $\alpha_0 = 1$ is the value of the bifurcation parameter before the start of the linear ramp, and $\Theta$ is again the Heaviside function. The red noise model underlying Fig. 2B uses the same noise term as in the above linear model but with $\kappa = 0.0045$, $\tau_0^{\text{noise}} = 1/4$ and $\tau_{t_c}^{\text{noise}} = 10$.

Instead of the fold bifurcation normal form, a specialized model might be more suitable to represent AMOC dynamics. To this end, we implement the dimensionless version of the 2D Stommel-Cessi model (*39*) given by

$$dX_t = b(-X_t(1 + \eta^2(X_t - Y_t)^2) + \alpha(t))dt + \sqrt{b}\sigma dW_t^X \quad (11)$$

$$dY_t = b(-\varepsilon^{-1}(Y_t - 1) - Y_t(1 + \eta^2(X_t - Y_t)^2))dt + \sqrt{b}\sigma dW_t^Y \quad (12)$$

with $b = 0.1$, $\varepsilon = 0.01$, $\sigma = 0.04$, and $\eta^2 = 7.5$. The two white noise terms acting on the components are independent. $\alpha(t)$ decreases from $\alpha_0 = 5$ to $\alpha_{t_c} = 1.128$. This decrease is linear or nonlinear for the analyses of Fig. 3 (A and B), respectively. All of the models were integrated using the Euler-Maruyama scheme to obtain time series data.

### Estimation methods for time of tipping

Three approaches to estimating the tipping time have been discussed and compared quantitatively in the Results section. We refer to them here as (i) AC(1) extrapolation, (ii) λ extrapolation, and (iii) the MLE method introduced by DD23. All of them build on the assumption that the system in question is well-represented by a 1D fold-type bifurcation in its normal form. The corresponding model equation is given by the deterministic part of Eq. 8. Methods 1 and 3 also assume the given stochastic white noise part of the same equation, while method 2 relies on a model suited for both white and nonstationary correlated (red) noise. We give here a more detailed description of the first two methods. For more information on the third method, we refer to (*14*), where it was originally proposed.

The synthetic time series underlying the comparison of the methods in the Results section consist of a stationary and a nonstationary part each. The MLE method uses both parts to fit a model of a linear control parameter ramp starting from a known time $t_0$. The extrapolation methods use only the nonstationary part after a known time $t_0$ as a basis for the linear fits.

1. AC(1) extrapolation: Before the annihilation of a system's equilibrium point, the negative feedbacks defining said equilibrium weaken with respect to the positive feedbacks. This results in a decrease of the restoring rate with respect to small noisy disturbances. If these disturbances are assumed to be stationary in time, the the lag-1 autocorrelation [AC(1)] can serve as a measure of system stability, as it can be shown to be a function of the linear restoring rate γ, i.e., the amplitude of the continuous linearized dynamics around the equilibrium point. In particular, when linearizing the dynmics under the assumption of white noise forcing, one arrives at the Ornstein-Uhlenbeck equation representing the evolution of small disturbances under a negative linear feedback or linear restoring rate, γ

$$dX_t = -\gamma X_t dt + \sigma dW_t \tag{13}$$

The discrete time approximation of these linearized dynamics is $X_{t+\Delta t} = \exp(-\gamma \Delta t)X_t + \eta_t$, where $\eta_t$ denotes the noise. The AC(1) of discrete samples of this stochastic process is thus given by AC(1) = $\exp(-\gamma \Delta t)$. Here, $\Delta t$ denotes the time step between subsequent measurements, which we set equal to one in our simulations. As we approach the tipping point $\gamma \to 0$, and thus, the AC(1) parameter approaches +1. Linearizing the fold-bifurcation normal form in Eq. 8, one sees that $\gamma = 2b\sqrt{\alpha}$, where $\alpha$ is the bifurcation parameter. Thus, $\alpha$ is approximately linearly related to $\log[AC(1)]^2$. Observing the latter quantity and extrapolating its best linear fit toward the bifurcation threshold, which for the fold normal form is at $\alpha_{tc} = 0$, yields an estimate of the tipping time.

2. $\lambda$ extrapolation: To account for possible nonstationary time correlation in the driving noise, Boers (8) estimated system stability via regressing the increments $X_{t+\Delta t} - X_t$ onto $X_t$ using a GLS method designed for models driven by red noise of varying correlation strength. This regression gives the autoregression parameter $\lambda$ in a model driven by discrete time red noise

$$X_{t+\Delta t} - X_t = \lambda X_t + \kappa \eta_t \tag{14}$$

$$\eta_{t+\Delta t} = \rho \eta_t + \epsilon_t \tag{15}$$

where $\rho$ symbolizes the correlation parameter of the red noise, and $\epsilon$ is the white noise. For correlation parameter $\rho$ close to zero, i.e., $\eta$ close to white noise, it can be shown that the relation $\lambda = AC(1) - 1 = \exp(-\gamma \Delta t) - 1$ holds approximately. Hence, as $\gamma \to 0$ from above, we have that $\lambda \to 0$ from below. Similarly to the above, the tipping time can thus be estimated by taking the time at which the linearly extrapolated $\log(\lambda + 1)^2$, as an estimate of the bifurcation parameter $\alpha$, crosses the critical value $\alpha = 0$. In the present applications, the quantities $\log[AC(1)]^2$ and $\log(\lambda + 1)^2$ are estimated in rolling windows centered around every 50th time series entry with a length of 100 time series entries each.

**Supplementary Materials**
**This PDF file includes:**
Figs. S1 and S2
Tables S1 and S2

**Acknowledgments**
**Funding:** M.B.-Y. and N.B. acknowledge funding by the European Union's Horizon 2020 research and innovation programme under the Marie Sklodowska-Curie grant agreement no. 956170. N.B. and S.B. acknowledge funding by the Volkswagen foundation. This is ClimTip contribution #1; the ClimTip project has received funding from the European Union's Horizon Europe research and innovation programme under grant agreement no. 101137601. **Author contributions:** M.B.-Y., A.M., S.B., and N.B. conceived and designed the study. A.M. carried out the analysis for Figs. 1, 2, and 3, and M.B.-Y. carried out the analysis for Figs. 4 and 5. All authors contributed to writing the manuscript. **Competing interests:** The authors declare that they have no competing interests. **Data and materials availability:** The HadISST1, HadSST4, and HadCRUT5 datasets are all available at https://metoffice.gov.uk/hadobs/. The ERSSTv5 operational data are available at https://psl.noaa.gov/data/gridded/data.noaa.ersst.v5.html. $CO_2$ emissions from the SSP scenarios can be found at https://tntcat.iiasa.ac.at/SspDb/ and historical $CO_2$ emission data at https://ourworldindata.org/co2-emissions. $CO_2$ emissions expected from current policies and targets can be found at https://climateactiontracker.org/. The generated synthetic time series and AMOC fingerprint time series can be found at https://zenodo.org/records/12549739. All other data needed to evaluate the conclusions in the paper are present in the paper and/or the Supplementary Materials. All code used to analyze the data and generate figures can be found at 10.5281/zenodo.12569006 (GitHub link: github.com/TUM-PIK-ESM/tipping_time).

Submitted 20 October 2023
Accepted 27 June 2024
Published 2 August 2024
10.1126/sciadv.adl4841

# Science Advances

## Supplementary Materials for

**Uncertainties too large to predict tipping times of major Earth system components from historical data**

Maya Ben-Yami *et al.*

Corresponding author: Maya Ben-Yami, maya.ben-yami@tum.de; Niklas Boers, boers@pik-potsdam.de

**This PDF file includes:**

Figs. S1 and S2
Tables S1 and S2
References

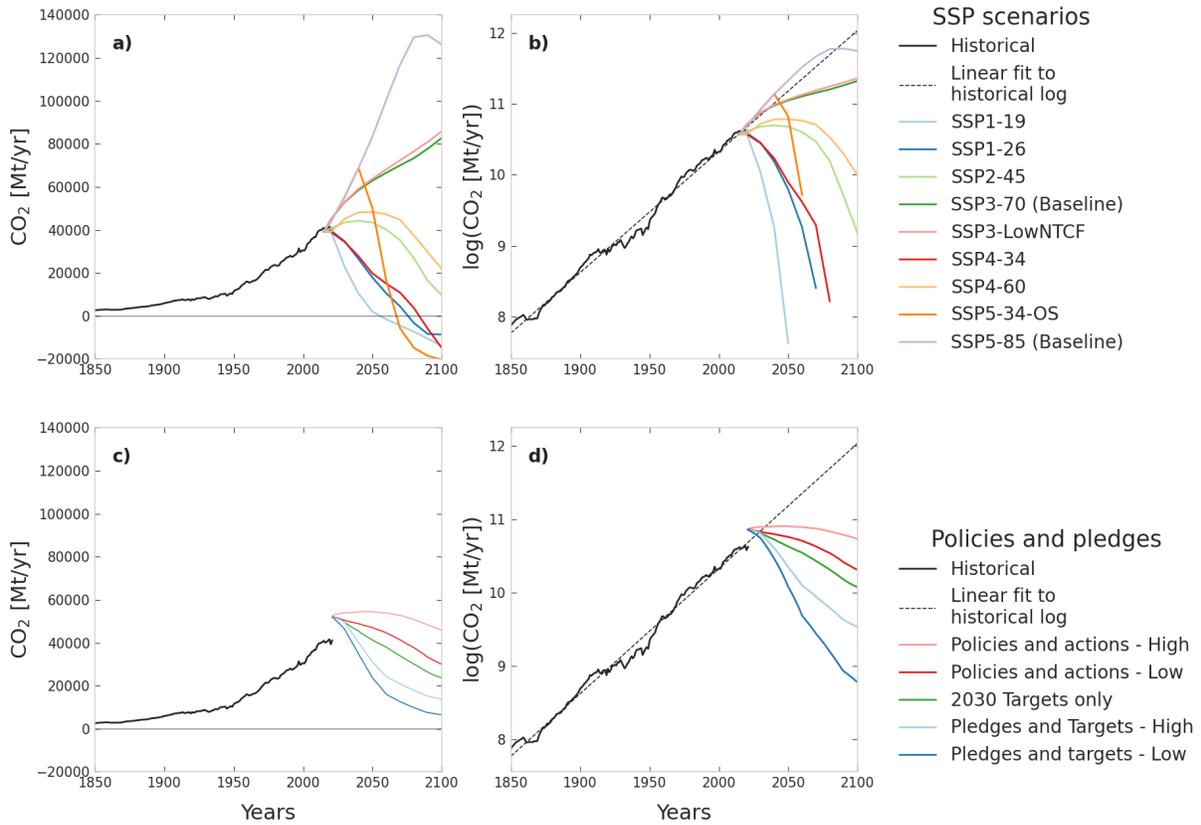

Figure S1: **Meaning of "business as usual"** Historical $CO_2$ emissions (black, a-d), SSP projected $CO_2$ emissions (colors, a-b)(*72*) and calculated emissions given current policies and actions (colors, c-d)(*73*). Both the raw emissions (a,c) and their natural logarithm (b,d) are shown. A linear fit to the historical log($CO_2$) data is also shown (dashed black line).

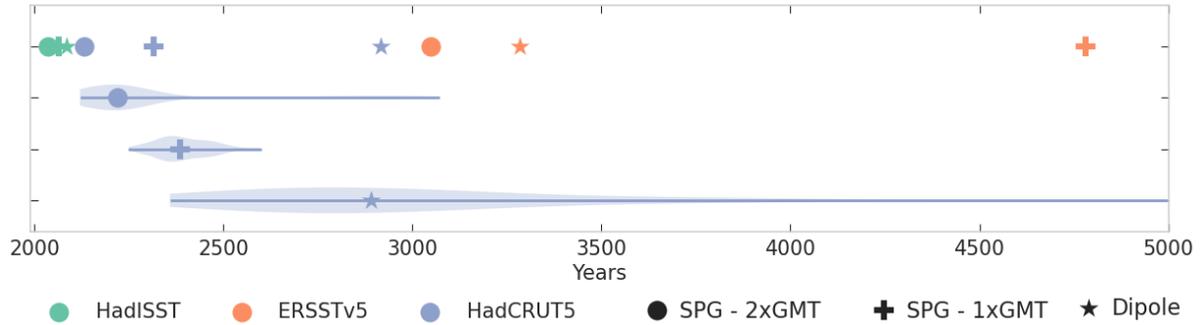

Figure S2: **Tipping times with penalization procedure from Ditlevsen & Ditlevsen 2023** (*14*). Same as Figure 5 with the penalization procedure from DD23 (cf. Section S2 of that paper). Note that out of the 200 members for the dipole fingerprint, in 47 members the tipping time went to infinity when the optimization was attempted, and they are not included here.

|            | p=0            | optimal p      |
|------------|----------------|----------------|
| **SPG-2xGMT** | 2158.3-3072.8  | 2120.9-3072.8  |
| **SPG-1xGMT** | 2249.8-2992.6  | 2249.8-2600.7  |
| **Dipole**    | 2428.2-8065.1  | 2359.1-inf     |

Table S1: AMOC minimum to maximum tipping times calculated using DD23's MLE method for the HadCRUT5 uncertainty ensemble, both without (p=0) and with (optimal p) the penalization procedure (cf. Section S2 of DD23 for details). The distribution can be seen in Figures 5 and S2.

|                     | optimal p | p=0    |
|---------------------|-----------|--------|
| SPG-2xGMT HadISST   | 2037.5    | 2053.7 |
| SPG-2xGMT ERSSTv5   | 3047.7    | 3047.7 |
| SPG-2xGMT HadCRUT5  | 2131.6    | 2168.9 |
| SPG-1xGMT HadISST   | 2064.7    | 2084.2 |
| SPG-1xGMT ERSSTv5   | 4780.9    | 4780.9 |
| SPG-1xGMT HadCRUT5  | 2314.6    | 2314.6 |
| Dipole HadISST      | 2085.2    | 2123.9 |
| Dipole ERSSTv5      | 3285.4    | 3285.4 |
| Dipole HadCRUT5     | 2915.5    | 2914.3 |

Table S2: Tipping times calculated using DD23's MLE method, both without (p=0) and with (optimal p) the penalization procedure (cf. Section S2 of DD23 for details). Note that for some of the larger tipping times the optimal p is 0, so the values are the same.